# Generation of narrow peaks in charged-particle magnetic spectroscopy

Dirk Dubbers

*Physikalisches Institut der Universität, Im Neuenheimer Feld 226, 69120 Heidelberg, Germany*
*E-mail address:* dubbers@physi.uni-heidelberg.de

14 October 2015

ABSTRACT

In spectroscopy of charged particles, narrow peaks may appear in initially smooth continuous spectra if magnetic transport of the particles is involved. As such unexpected peaks may be misinterpreted as new physics, their generation is investigated for various experimental configurations.

## 1. Phenomenology

New particles are often discovered as unexpected peaks in otherwise smooth spectra. Therefore it is of general interest to identify sources of artificial peaks that may lead to wrong conclusions. The purpose of this letter is to show how easy it is to produce narrow peaks in seemingly well understood instrumental configurations.

Let us begin with a most simple experiment. Place a small isotropic source of the $\beta$-emitter $^{90}$Sr-$^{90}$Y in a uniform magnetic field, applied along axis $z$. The two successive $\beta$-transitions have endpoint energies $E_0 = 0.55$ and 2.28 MeV. In a guide field of $B_0 = 0.5$ T, radii of electron gyration at midpoint of the two $\beta$-spectra are $r_0(E_0/2) = 1.0$ cm and 0.4 cm, under emission perpendicular to the field.

In the uniform field, we install a circular electron detector at a distance $z_0 = 0.2$ m to the source, with radius $r_2 = 2$ cm. This is large enough to contain all electrons, because $2r_0(E_0) = 1.9$ cm. Curve a) in Fig. 1 shows the $\beta$-spectrum seen in the detector. To suppress the low-energy $^{90}$Sr spectrum, we cover the center of the detector with a small electron absorber of radius $r_1 = 0.5$ cm. Curve b) of Fig. 1 shows the calculated spectrum for this case. Surprisingly, the suppression of low energy $^{90}$Sr $\beta$-particles leads to a modulation in the high-energy part of the $^{90}$Y spectrum.

If we remove the absorber and place both source and detector into the non-uniform field at the ends of a 30 cm long solenoid (insert to Fig. 1), where the field $B = 0.25$ T has dropped to one half, spectrum c) appears. The peaks sharpen further, spectrum d), when the size of the detector is reduced to $r_2 = 1$ cm. The dotted curve shows what conventional theory would predict for this case. The absolute heights of the spectra are known, but are adapted as to fit best into the plot. The spectra shown for an energy resolution of 20 keV.

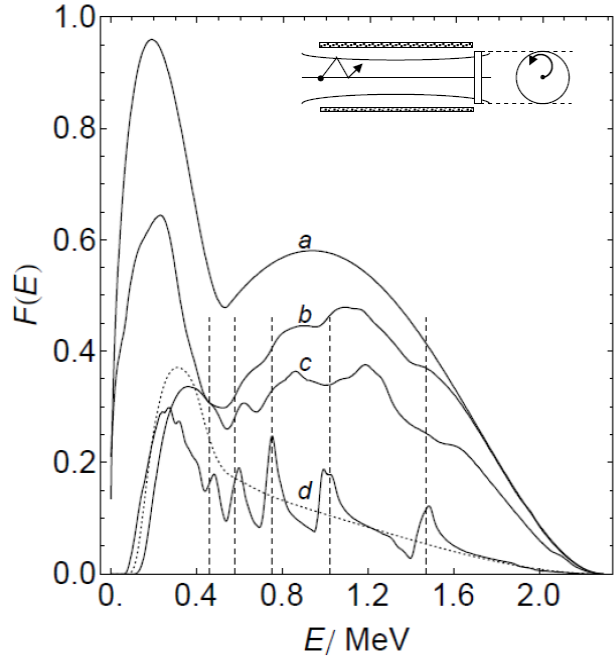

**Fig. 1.** Development of peaks from an initially smooth spectrum: **(a)** β-spectrum from $^{90}$Sr-decay. **(b)** Spectrum after electron transport in a uniform guide field to a detector, whose inner 6% area is covered by an electron absorber. **(c)** Source and detector (without absorber) are placed in the fringe field at the ends of a solenoid (see insert, side view and axial view). **(d)** Like (c), but with a reduced detector diameter, and with the absorber of b) reinstalled, calculated with the true PSF (full line) and the conventional PSF (dotted line). The vertical dashed lines are discussed later in the text (cf. Fig. 5).

The following Sect. 2 summarizes recent results on the so-called magnetic point spread function $f(R)$ (PSF) and its singularities. Section 3 explains the peaks seen in the spectra $F(E)$ of Fig. 1. Sections 3 and 4 discuss setups where even stronger peaks appear, and present results in the context of earlier heavy-ion experiments that had used magnetic transport in positron spectroscopy.

## 2. The magnetic PSF for monoenergetic electrons and its caustics

Recently spectroscopy of charged particles guided by a magnetic field was investigated [1], and it was shown that, due to caustic effects, an infinite number of singularities appear in the point spread function. These singularities, studied in more detail in Ref. [2], apparently had so far escaped the attention of investigators. In the meantime, the new effect was verified experimentally at Los Alamos [3], and further investigated in Ref. [4]. These singularities may play a role beyond nuclear and particle spectroscopy, in topics listed in [2], like surface photoelectron spectroscopy, reaction microscopes, or retardation spectroscopy.

In Ref. [2] it was pointed out that the positions $R_n$ of the singularities in $f(R)$ are extremely sensitive to parameters like particle energy, and it was argued that, for a continuous energy spectrum, the effects of these caustics would rapidly average out. However, the present paper presents a number of examples where this expectation is deceived.

An electron (or other particle of charge $e$ and mass $m$) emitted from a point source with kinetic energy $E$, under polar angle $\theta$ relative to field axis $z$, has a gyration radius $r = r_0 \sin\theta$, with $r_0 = p/eB$ and relativistic momentum $p = c^{-1}E^{1/2}(E+2mc^2)^{1/2}$. After magnetic transport in the uniform field, the electron reaches the detector at distance $z_0$, after a number of $\alpha/2\pi = z_0/d$ gyration orbits, with helix pitch $d = 2\pi r_0 \cos\theta$. When the electron hits the detector, its phase angle $\alpha$ there is predetermined by the emission angle $\theta$ at the source,

$$\alpha = z_0 / (r_0 \cos\theta). \tag{1}$$

The point of impact of the electron on the detector surface is displaced from the central point of impact, reached under $\theta = 0$, by a distance

$$R(\alpha) = 2r|\sin \tfrac{1}{2}\alpha| = 2r_0\sqrt{1-\alpha_0^2/\alpha^2}\,|\sin \tfrac{1}{2}\alpha|, \tag{2}$$

as follows from simple geometry. The smallest phase angle $\alpha_0 = z_0/r_0$ is for emission in the limit $\theta \to 0$. The corresponding minimum number of orbits $n_0 = \alpha_0/2\pi$, or

$$n_0 = eBz_0/2\pi p, \tag{3}$$

is the only instrument parameter of the theory. For a non-uniform field, $Bz_0$ must be replaced by the corresponding field integral.

The function $R(\alpha)$ is shown in Fig. 2a for $n_0 = 8.4$ gyration orbits. $R$ is zero for $\alpha < \alpha_0$, is zero again for every full number $n$ of orbits, and attains maxima $R_n$ in between, near every half-integer number $n+½$ of orbits. Figure 2b shows the same function in polar coordinates in the $x$-$y$ plane at the $z = z_0$.

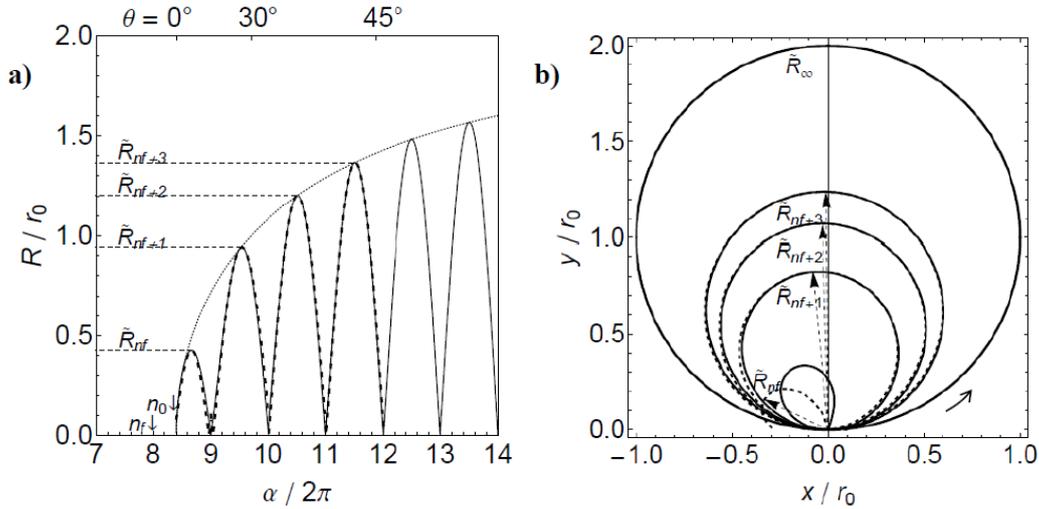

**Fig. 2.** (a) Electron displacement $R$ on the detector, Eq. (2), as a function of phase angle $\alpha$, for instrument parameter $n_0 = \alpha_0/2\pi = 8.4$ orbits. The dashed curves are our approximation to $R(\alpha)$, Eqs. (6) and (7), with $n_f = \text{floor}(n_0)$, and $\tilde{R} = R/r_0$. The corresponding emission angles from Eq. (1) are given on the upper axis. (b) The same function in polar coordinates, for an electron starting at the source with azimuthal angle $\varphi = 0$.

For a given azimuthal starting angle $\varphi$, an electron emitted with polar angle $\theta$ reaches one single point on the detector, whose phase angle $\alpha$ is fixed by Eq. (1). Different values of $\alpha$ are reached only upon a change of $\theta$, and with it of gyration radius $r = r_0 \sin\theta$. With increasing $\theta$, the points of arrival on the detector therefore spiral successively through near-circular curves, as shown in Fig. 2b. These trajectories then must be averaged over $\varphi$. A singularity appears in the PSF whenever the displacement $R$ is stationary. The effect is related to the development of caustics (or of density-of-state singularities) in other fields of physics.

The magnetic PSF is the radially symmetric distribution function $f(R) = 1/(2\pi R) \, dP/dR$ of the electrons on the detector, where $dP$ gives the probability for finding the particle displaced by a distance between $R$ and $R+dR$ from its initial field line. For our purposes, the following presentation of the PSF is useful

$$f(R) = \frac{1}{2\pi R} \frac{dP}{d\cos\theta} \frac{d\cos\theta}{d\alpha} \frac{d\alpha}{dR}, \quad (4)$$

with $dP/d\cos\theta = 1$ for isotropic particle emission. The derivative $d\alpha/dR$ therein produces a singularity whenever $R(\alpha)$ reaches a maximum $R_n$. From Eqs. (1), (2), and (4), the PSF is obtained as a function of $\alpha$,

$$f(\alpha) = \frac{1}{2\pi R r_0} \frac{\alpha_0 (\alpha^2 - \alpha_0^2)^{1/2}}{\left| \alpha(\alpha^2 - \alpha_0^2)\cos(\alpha/2) + 2\alpha_0^2 \sin(\alpha/2) \right|}, \quad (5)$$

with $R$ from Eq. (2), valid from $\alpha = \alpha_0$ for emission under $\theta = 0$ up to $\alpha \to \infty$ for $\theta = \pi/2$.

However, what we need is not $f$ as a function of $\alpha$, but the PSF $f(R)$ as a function of $R$, and we must find the inverse function $\alpha(R)$ of Eq. (2). This cannot be done algebraically, and some approximation is required. The function $R(\alpha)$ can be made invertible, separately for each full orbit numbered $n$, by approximating it by cosine functions that pass through the maxima $R_n(\alpha_n)$. For the lowest orbit numbered $n = n_f$, this gives

$$R(\alpha) \approx R_{n_f} \cos \frac{\alpha - \alpha_{n_f}}{2(n_f + 1 - n_0)}, \quad (6)$$

where $n_f = \text{floor}(n_0)$ is the next lower integer to $n_0$. For the subsequent orbits numbered $n > n_f$,

$$R(\alpha) \approx R_n \cos[(\alpha - \alpha_n)/2], \quad (7)$$

where the $\alpha_n$, $n = n_f, n_f +1, \ldots$, are the positions of the maxima $R_n$ on the abscissa of Fig. 2a.

The inverted functions, to be inserted into Eq. (5), then are, for the lowest and following orbits,

$$\alpha_{n_f \pm}(R) \approx \alpha_{n_f} \mp 2(n_f + 1 - n_0) \arccos(R/R_{n_f}), \quad (8)$$

$$\alpha_{n\pm}(R) \approx \alpha_n \mp 2\arccos(R/R_n), \text{ for } n > n_f, \quad (9)$$

with the minus sign for the rising branches of $R(\alpha)$ in Fig. 2, the plus sign for the falling branches. In this way one obtains from Eq. (5) a partial PSF $f_n = f_{n+} + f_{n-}$ for every orbit $n$. Each $f_n(R)$ ends with a singularity at $R = R_n$, whose shape is largely independent of the shape of the other singularities. All $f_n$ must then be added up to obtain the full PSF $f(R)$. Note that no separate integration over polar emission angle $\theta$ is required, because $\theta$ is linked to $\alpha$ via Eq. (1).

The true orbits $R(\alpha)$ (full curves in Fig. 2) and the approximate orbits (dashed curves) are both known analytical functions. While the approximate curves give the correct positions $R_n$ of the singularities, the azimuthal angles in the *x-y* plane, Fig. 2b, show stronger deviations, in particular for the lowest orbit. These deviations can be suppressed by slightly varying one so far unused parameter in the approximation, namely, the widths of the cosines in Eqs. (6) and (7). I then recalculated the PSFs $f(R)$ and the spectra $F(E)$, using these adapted widths, and found that changes were hardly visible. From this I conclude that the deviations seen in Fig. 2 are innocent for our purposes.

## 3. Peaks in continuous spectra for rotationally symmetric detection

What do these results for the monoenergetic PSF $f(R)$ mean for particles with a continuous energy spectrum $F(E)$? We discuss this first for the examples of Sect. 1. For rotationally symmetric detectors, the function $g(R) = 2\pi R f(R)$ gives the probability that the electron hits the detector anywhere at a distance $R$ from the origin. For a given energy $E$; this response is calculated from Eqs. (5) through (9), and the detector count rate is obtained by integration of $g(R)$ over $R$ from the radius $r_1$ of an absorber ($r_1 = 0$ if there is none) to the radius $r_2$ of the detector.

The resulting peaks in Fig. 1 can be understood by looking at the ring-shaped PSF on the detector surface, shown in Fig. 3 for monoenergetic electrons, with $n_0 = 8.4$ ($n_f = 8$). When the energy $E$ of the incident electrons is increased, more and more ring shaped singularities dwell out of the origin $R = 0$, as shown in Fig. 4, which displays the dependence of the radii $R_n$ on electron energy $E$.

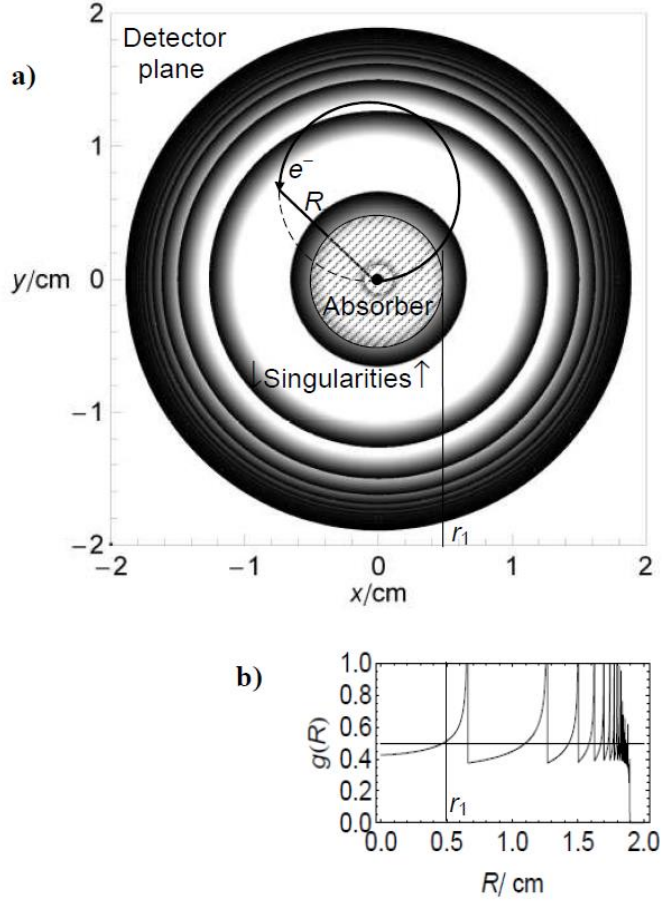

**Fig. 3. (a)** The ring-shaped singularities in the PSF $f[(x^2+y^2)^{1/2}]$, as they appear on the surface of the detector of Sect. 1(plus absorber), for energy $E = 1$ MeV and a field $B = 0.5$ T at right angle to the paper plane. A singularity with $n_f = 3$ has just crossed the edge $r_1$ of the absorber. The central dot is the source at $z_0 = 0.2$ m distance. An electron trajectory is shown for $\theta = 45°$, reaching the detector after 4.7 orbits. **(b)** Probability $g(R) = 2\pi R\,f(R)$ for finding the electron at displacement $R$.

A peak is expected in the spectrum whenever a new singularity in Fig. 3 has entered the active area of the detector at $R_n(E) = r_1$. Putting the maxima $R_n$ in Fig. 2 in the middle of the $n^{th}$ interval (Fig. 2), we find the energies where this happens as

$$E_{cn} = \sqrt{b^2\,[\pi^2 r_1^2 + z_0^2/(n+1/2)^2] + m^2 c^4} - mc^2, \qquad (10)$$

with field parameter $b = ceB/2\pi$. The $E_{cn}$ are rather precise for $n > n_f$, but less so for values of $n_0$ near 1, found at high energy or low field. Setting $r_1 = 0$ in Eq. (10) gives the energies $E_{0n}$ at which a new singularity comes out of the origin $R = 0$, for integer $n_0$. From Eq. (3) we see that, with increasing $E$, the minimum number $n_0$ of orbits between source and detector decreases, until there is no longer a full orbit under $\theta = 0$.

In Fig. 4 the energies $E_{0n}$ where $R_n(E) = 0$ are indicated by dots on the abscissa, and the energies $E_{cn}$ by vertical dashed lines and dots at $R_n(E) = r_1$. The parameters used in Fig. 4 are those of the experiment EPOS II, to be treated in a later section. The rightmost curve starts at $n_f = 6$, the curves to its left have $n_f = 7, 8, 9, \ldots$

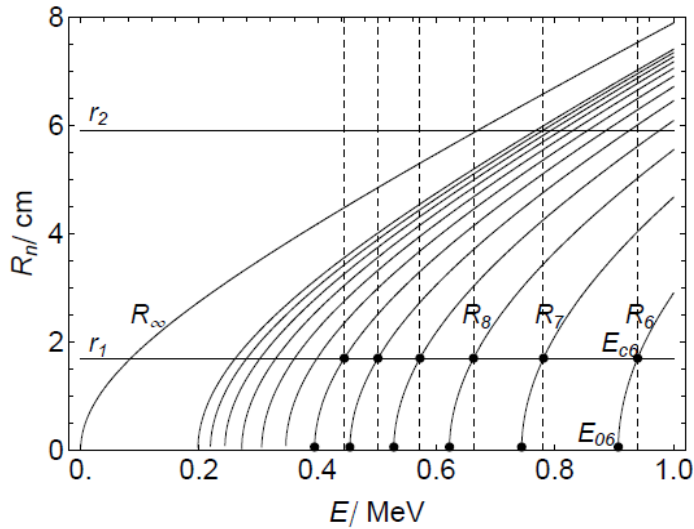

**Fig. 4.** Radii $R_n$ of the ring shaped singularities of the PSF as functions of particle energy $E$. Whenever $E$ crosses one of the energies $E_{0n}$ (dots on the abscissa), a new singularity dwells out of the center $R = 0$. With $E$ growing further to energy $E_{cn}$ from Eq. (10) (vertical dashed lines), the new singularity reaches the active area of the detector at $R_n = r_1$. The figure is calculated for the parameters of EPOS II, Sect. 4.

With no absorber and for a large detector in a uniform magnetic field, no peaks should appear in the spectrum, because the PSF for isotropic emission is normalized to one [5], and hence integration over $R$ from zero to infinity gives unity for any energy. However, if the center of the detector is blocked by an absorber, detector response changes abruptly when a new singularity enters the active area, i.e., when its radius $R_n$ grows beyond absorber radius $r_1$, as seen in curve b) of Fig. 1. Alternatively, instead of using an absorber, we can disturb isotropy by using the magnetic mirror (insert to Fig. 1). For a field $B_0$ in the center of the solenoid, and a field $B < B_0$ at the source, the critical angle of mirror reflection back to the source is $\theta_c = \arcsin\sqrt{B/B_0}$, and electrons emitted with $\theta > \theta_c$ do not reach the detector. Curve c) shows the resulting spectrum for $B = B_0/2$, with $\theta_c = 45°$. The peaks sharpen further, curve d) in Fig. 1, if the absorber is reinstalled and the radius of the detector halved to $r_2 = 1$ cm. Conventional theory, which does not know caustics (dotted line in Fig. 1), was summarized in Sect. 2 of Ref. [1].

Next we apply our algorithm to the detector setup used in an earlier experiment called EPOS, done at the UNILAC accelerator of GSI. In this experiment, beams of heavy ions of combined nuclear charge near $Z = 180$ were colliding head-on, with energies of several MeV per nucleon. In the collision, positrons with a continuous spectrum were created in various nuclear and atomic processes. To decrease background, the positrons were magnetically guided to a distant detector. The initial aim of the experiments was to search for spontaneous $e^+$- $e^-$ pair creation, signaling the breakdown of the vacuum in super-intense fields.

In the first experiments EPOS I, Ref. [6] and references therein, significant and unexpected peaks in the measured positron spectra had aroused considerable interest. Our caustic-induced peaks have widths and separations similar to those seen in these early heavy-ion experiments, and I want to investigate whether both phenomena are related to each other.

During magnetic transport, the helical trajectories of the positrons periodically cross the $z$-axis on which the source is positioned. Therefore a "pencil detector" had been chosen in EPOS I, with the shape of a narrow cylinder of radius $r_2 = 0.5$ cm and length $l = 8$ cm, installed at $z_0 = 0.83$ m, oriented along $z$. The active detector area was the mantle of this cylinder (see insert to Fig. 5). For a guide field $B = 0.18$ T, the detector will intercept all electrons, emitted under $\theta > 45°$, whose pitch obeys $2\pi r_0 \cos\theta < l$, or whose energy is below 0.6 MeV, which covers most of the spectrum in Fig. 5.

The detector hence measured the positrons at a displacement fixed to $R = r_2 = 0.5$ cm, with no integration over $R$ needed. With increasing energy $E$, successive singularities cross the radius $r_2$ of the detector. At a given distance $z$, the singularities occur for crazing incidence of the positrons on the detector, and one might suspect that they are suppressed by the finite dead layer of the detector. However, this poses hardly a problem because the widths of the resonances in $f(R)$ are of order several millimeter (like in Figs. 3b and 6b), while detector radius is $r_2 = 5$ mm. I recall that a given $g(r_2)$, calculated from Eqs. (5) to (9), sums up all angles of incidence on the detector surface, be $g(r_2)$ on resonance, $r_2 = R_n$ or not.

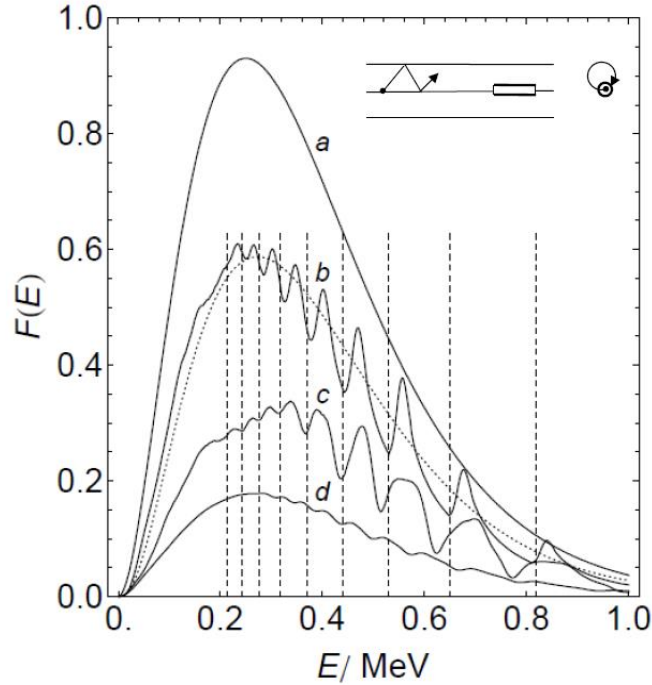

**Fig. 5.** Positrons, transported to a pencil detector (see insert, side view and axial view), have a peaked spectrum when detected under $2\pi$. The peaks disappear when low emission angles are suppressed, as it happened in EPOS I: **(a)** The initial spectrum, as generated in heavy ion collisions. **(b)** Response in the limit of zero detector length, calculated both with the true PSF (full line) and the conventional PSF (dotted line). **(c)** Response for the detector of full length. **(d)** Response for emission angles limited, as in EPOS I, to $45° < \theta < 90°$. The vertical dashed lines are the EPOS I equivalent to the lines in Fig. 4.

The extension of the EPOS I detector surface into the $z$-direction has not a big effect on the spectrum, because detector length $l = 8$ cm $\equiv \Delta z_0$ is small compared to detector distance, leading merely to small shifts $\Delta n_0/n_0 = \Delta z_0/z_0$ in the instrument parameter $n_0$ from Eq. (3).

During integration over detector length $l$, care must be taken that a positron is registered only at its first encounter with the detector.

Figure 5 shows the calculated detector response for the EPOS I setup, with the source assumed to be isotropic. Curve a) shows the initial undisturbed positron spectrum, adapted from Fig. 10 of Ref. [7]. Spectrum b) gives the response of the pencil detector, calculated in the limit of zero length. Curve c) shows the same spectrum, integrated over the 8 cm length of the detector. Spectrum d) shows the response if the positrons reaching the detector are limited to large polar emission angles $\theta > 45°$, as it was the case in EPOS I. The rightmost vertical dashed line in Fig. 5 has $n_f = 4$, followed by $n_f = 5, 6, \ldots$ for the lines to its left.

In the EPOS I experiment, the suppression of angles $\theta < 45°$ was due to a propeller-like beam element that prevented counter-circulating electrons from reaching the detector. With only large emission angles reaching the detector, the peaks are almost completely suppressed, and therefore are not thought to be at the origin of the peaks seen in EPOS I. Had the angle-limiting beam elements been removed during measurements with a $\beta^+$ test source, then the caustics-induced peaks should have become visible in EPOS I.

The APEX experiment, done at Argonne's ATLAS heavy-ion accelerator, pursued the same questions as EPOS. It used a large pencil detector of radius $r_2 = 1.5$ cm and length $l = 33$ cm, in a low guide field $B = 0.03$ T [8]. Like EPOS II (see below), it simultaneously measured positron and electron spectra with detectors at distances $z_0 = +1.5$ m and $z_0 = -1.5$ m to the source. Unlike EPOS, the experiment saw no peaks, even at considerably improved statistics. In APEX, glancing incidence of electrons and positrons with $R$ near $r_2$ was suppressed due to the pagode-like structure of its detectors. Therefore peaks due to caustics could not develop. The structure of the APEX detectors is difficult to implement in our code, and we do not further investigate its response.

## 4. Non-rotational symmetric detection

In the case of non-rotational detectors symmetry, the distribution $g(R)$ must be replaced by the (equally rotational symmetric) PSF $f[(x^2+y^2)^{1/2}]$, to be integrated over the sensitive surface of the detector. This case applies to the improved GSI instrument EPOS II, which in the end also got rid of the peaks seen in EPOS I, see [9] and references therein.

The planar SiLi detector configuration of EPOS II, shown in Fig. 6, was installed at both ends $z_0 = \pm 1.5$ m of the instrument, in a guide field $B = 1.2$ T. Each detector had an active area extending from $x_1 = 1.7$ cm to $x_2 = 5.9$ cm and over length $l = 6$ cm along $z$. The central opening between the detectors allowed the huge number of so-called $\delta$-electrons pass the detectors incognito.

The dotted curve in Fig. 6b shows the PSF $f(x, y_0 = 0)$ for $E = 0.51$ MeV. The range $\Delta\alpha$ of phase angles accepted by the detectors varies with $R$. When, with increasing $E$, the first positrons hit the detector at $R = r_1$, they do this with $\alpha = \pi$ and zero range $\Delta\alpha$. When $R$ increases beyond $x_1$, the accepted range grows as

$$\Delta\alpha = \arcsin(1 - x_1^2 / R^2)^{1/2} \equiv \Delta\alpha_1, \text{ for } x_1 \leq R < x_2, \text{ and} \quad (11)$$

$$\Delta\alpha = \Delta\alpha_1 - \arcsin(1 - x_2^2 / R^2)^{1/2}, \text{ for } x_2 \leq R < 2r_0. \quad (12)$$

The width $\Delta\alpha/2\pi$ is also plotted in Fig. 6b, with $\Delta\alpha$ reaching up to 20% of $2\pi$. For the case (unrealized in EPOS) that emission angles are limited to $0° \leq \theta \leq 45°$, the distribution $\Delta\alpha \times f(x, y = 0)$ on the detector is also shown.

The spectrum $F(E)$ measured by the detectors is the integral of $\Delta\alpha \times f(x, y = 0, E)$ from $x_1$ to $x_1$, which can be further integrated over detector length $l$. This spectrum, calculated for the range $30° \leq \theta \leq 87°$ accessible in EPOS II, is shown in curve a) of Fig. 7, and no peaks are visible. Peaks due to caustics develop only for low emission angles, for instance, when angles $\theta > 45°$ (curve b) or $\theta > 35°$ (curve c) are excluded, which, however, was not feasible with EPOS II.

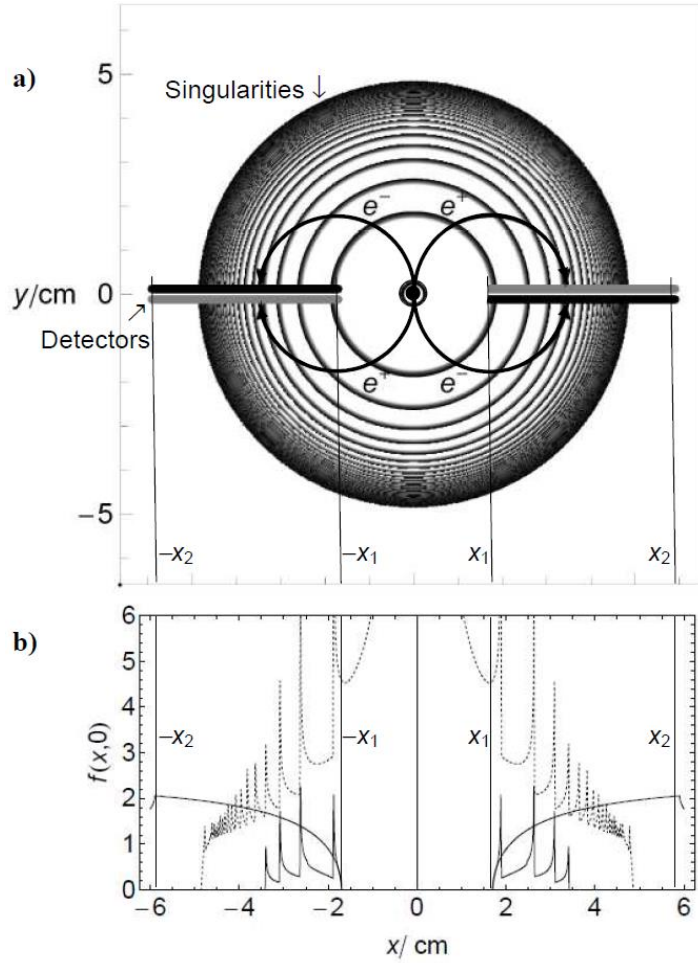

**Fig. 6. (a)** The ring-shaped PSF $f[(x^2+y^2)^{1/2}]$ at the detector site of EPOS II, for energy $E = 0.51$ MeV and a field $B = 0.12$ T at right angle to the paper plane. A singularity with $n_f = 6$ has just crossed the border $r_1$ of the detector, and a new singularity with $n_f = 5$ has just left the center. The central dot is the source at $z_0 = 1.5$ m distance. Trajectories of $e^+$ and $e^-$ are shown for $\theta = 43°$, starting at $\varphi = 0$ and reaching the detector after 13.5 orbits. **(b)** The PSF $f(x, 0)$ (dotted curve), the detector acceptance curve (full smooth curve) from Eqs. (11) and (12), and the distribution along the detector surface for an angular range limited to $0° \leq \theta \leq 45°$ (full peaked curve).

The methods described in this paper can be applied to arbitrary numbers and shapes of detectors. The complicated ($\theta$, $\varphi$) transmission plots for magnetic transport (see, for instance, Fig. 3.2 of Ref. [10]) are automatically respected when integration is restricted to the surface of the detector. When circular baffles, narrow vacuum tubes, or magnetic mirror effects limit emission angles $\theta$, then this can also be included in the code. The respective lower limiting values $\theta_1$, $\theta_2$, … and/or upper limiting values $\theta_1$', $\theta_2$', … then are translated via Eq. (1) into corresponding limits on $\alpha$, used in Eqs. (8) and (9),

$$\alpha_{min} = \max(\alpha_1, \alpha_2, ...), \quad \alpha_{max} = \min(\alpha'_1, \alpha'_2, ...), \tag{13}$$

which usually are energy dependent, except for magnetic mirrors. Finally, finite source volumes can be examined by varying $x_1$ and $z_0$. If more complicated structures are inserted into the beam line, or if coincidences with other particles (gamma rays from $e^+$ annihilation, etc.) are required, then one must rely on the usual Monte Carlon simulations of the setup.

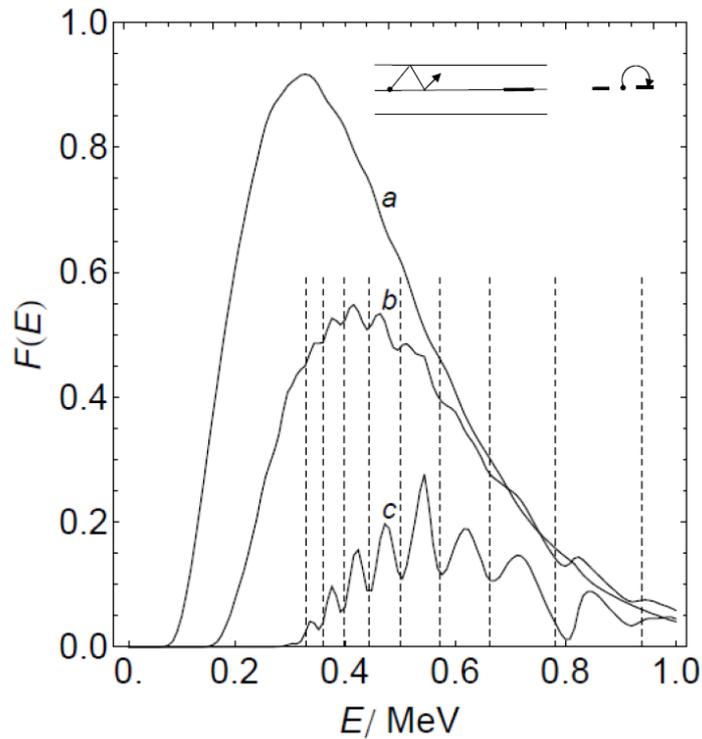

**Fig. 7.** Positrons, transported to one of the flat detectors of EPOS II (see insert, side view and axial view), have a smooth spectrum when detected under $2\pi$ (unlike EPOS I, Fig. 5b). Peaks would appear only if high emission angles were suppressed: **(a)** Calculated response of the EPOS II detectors, with its $30° \leq \theta \leq 87°$ acceptance. **(b)** Expected response of the EPOS II detectors if high-angle emission $\theta > 45°$ had been suppressed. **(c)** Same as b), for $\theta > 35°$ suppression. The vertical dashed lines are the same as in Fig. 4.

## Conclusions

The caustic-like singularities found recently in the point spread function $f(R)$ for magnetic transport of monoenergetic electrons to their detectors, can, quite unexpectedly, lead to narrow peaks in continuous energy spectra $F(E)$. The generation of these peaks was investigated for various detector configurations. In some setups, narrow peaks are seen for full $2\pi$ detector acceptance. In other setups, peaks appear only under low-angles. The strange peaks seen in early heavy-ion induced $e^+$-$e^-$ spectroscopy were not caustics-induced, although the detectors used there were in principle sensitive to such peaks.

## Acknowledgements

This work was supported by the Priority Programme SPP 1491 of Deutsche Forschungsgemeinschaft.